\documentclass[conference,10pt,bottom=0.95in]{IEEEtran}
\IEEEoverridecommandlockouts
\usepackage{graphicx,color}
\usepackage{algorithmic}
\usepackage{lipsum}
\usepackage{caption} 
\usepackage{subcaption} 
\usepackage{subfig}

\usepackage{amsmath, amsthm, amsfonts, amssymb, amsbsy,nccmath}
\usepackage{algorithm}
\usepackage{enumerate}
\usepackage{lipsum}

\usepackage{textgreek}
\usepackage{comment}
\usepackage{textcomp}
\usepackage{amsfonts}

\usepackage[sort,compress]{cite}
\usepackage{epsfig}
\usepackage{epstopdf}
\usepackage{mathtools}
\usepackage{dsfont}
\usepackage{cleveref}
\usepackage{subfig}
\usepackage{stfloats}
\usepackage{bbm}

\usepackage{epstopdf}

\theoremstyle{boldthm}
\newtheorem{definition}{Definition}[section]

\usepackage[inline]{enumitem}   
\makeatletter
\newcommand{\inlineitem}[1][]{%
\ifnum\enit@type=\tw@
    {\descriptionlabel{#1}}
  \hspace{\labelsep}%
\else
  \ifnum\enit@type=\z@
       \refstepcounter{\@listctr}\fi
    \quad\@itemlabel\hspace{\labelsep}%
\fi}
\makeatother
\newtheorem{theorem}{Theorem}

\newcommand{\beq}{\begin{equation}}
\newcommand{\eeq}{\end{equation}}

\def\adots{\mathinner{\mskip0mu\raise0pt\vbox{\kern7pt\hbox{.}}\mskip3mu
          \raise4pt\hbox{.}\mskip3mu\raise8pt\hbox{.}\mskip0mu}}

\newcommand{\bmc}{{\boldsymbol c}}

\newcommand{\bmh}{\bfh}

\usepackage{bm}

\newcommand{\bme}{{\bm e}}
\newcommand{\bmx}{{\bm x}}

\renewcommand{\bmh}{{\bm h}}

\newcommand{\mS}{\mathcal{S}}

\newcommand{\mD}{\mathcal{D}}

\newcommand{\mA}{\mathcal{A}}

\newcommand{\mV}{\mathcal{V}}

\newcommand{\mM}{\mathcal{M}}

\newcommand{\bmm}{{\boldsymbol {m}}}
\newcommand{\bmo}{{\boldsymbol {o}}}

\newcommand{\bmn}{{\bm n}}
\newcommand{\bma}{{\bm a}}

\newcommand{\bmd}{\bm d}

\newcommand{\bit}{\begin{itemize}}
\newcommand{\eit}{\end{itemize}}

\newcommand{\mL}{\mathcal{L}}
\newcommand{\mG}{\mathcal{G}}

\renewcommand{\bmh}{{\boldsymbol h}}
\renewcommand{\bmh}{{\boldsymbol h}}

\DeclarePairedDelimiter\abs{\lvert}{\rvert}%

\usepackage{lipsum}

\usepackage{float}
\floatstyle{plaintop}
\restylefloat{table}

\makeatletter
\newcommand\fs@spaceruled{\def\@fs@cfont{\bfseries}\let\@fs@capt\floatc@ruled
  \def\@fs@pre{\vspace{0.5\baselineskip}\hrule height.8pt depth0pt \kern2pt}%
  \def\@fs@post{\kern1pt\hrule\relax}%
  \def\@fs@mid{\kern2pt\hrule\kern2pt}%
  \let\@fs@iftopcapt\iftrue}
  \newcommand\fs@betterruled{%
  \def\@fs@cfont{\bfseries}\let\@fs@capt\floatc@ruled
  \def\@fs@pre{\vspace*{5pt}\hrule height.8pt depth0pt \kern2pt}%
  \def\@fs@post{\kern2pt\hrule\relax}%
  \def\@fs@mid{\kern2pt\hrule\kern2pt}%
  \let\@fs@iftopcapt\iftrue}
\floatstyle{betterruled}
\restylefloat{algorithm}
\makeatother

\usepackage{subcaption} 

\usepackage{footnote}

\begin{document}
\title{Causal Model-Based Reinforcement Learning for Sample-Efficient IoT Channel Access\vspace{-2mm}}
\vspace{-4mm}
\author{Aswin Arun\IEEEauthorrefmark{1}, Christo Kurisummoottil Thomas\IEEEauthorrefmark{2}, Rimalpudi Sarvendranath\IEEEauthorrefmark{1}, and Walid Saad \IEEEauthorrefmark{3}\\  \IEEEauthorrefmark{1} Department of Electrical Engineering, Indian Institute of Technology, Tirupati, Andhra Pradesh, India.\\  \IEEEauthorrefmark{2} Department of Electrical and Computer Engineering, Worcester Polytechnic Institute, Worcester, MA, USA.\\
\IEEEauthorrefmark{2} Department of Electrical and Computer Engineering, Virginia Tech, Alexandria, VA, USA.\\
Emails:\{ee22b055, sarvendranath\}@iittp.ac.in, cthomas2@wpi.edu, walids@vt.edu.\vspace{-0.4cm}}
\maketitle
\vspace{-4mm}
\begin{abstract}
    Despite the advantages of multi-agent reinforcement learning (MARL) for wireless use case such as medium access control (MAC), their real-world deployment in Internet of Things (IoT) is hindered by their sample inefficiency. To alleviate this challenge, one can leverage model-based reinforcement learning (MBRL) solutions, however, conventional MBRL approaches rely on black-box models that are not interpretable and cannot reason. In contrast, in this paper, a novel causal model-based MARL framework is developed by leveraging tools from causal learning. In particular, the proposed model can explicitly represent causal dependencies
between network variables using structural causal models
(SCMs) and attention-based inference networks. Interpretable causal models are then developed to capture how MAC control
messages influence observations, how transmission actions
determine outcomes, and how channel observations affect
rewards. Data augmentation techniques are then used to generate synthetic rollouts using the learned
causal model for policy optimization via proximal policy
optimization (PPO). Analytical results demonstrate exponential sample complexity gains of causal MBRL over black-box approaches.  Extensive simulations demonstrate that, on average, the proposed approach can reduce environment interactions by $58\%$, and yield faster
convergence compared to model-free baselines. The proposed approach
inherently is also shown to provide interpretable scheduling decisions via
attention-based causal attribution, revealing which network
conditions drive the policy. The resulting combination of sample
efficiency and interpretability establishes causal MBRL as a
practical approach for resource-constrained wireless systems.
\end{abstract}

\vspace{-2mm}\section{Introduction}
\vspace{-1mm}

The proliferation of wireless sensor networks, internet of things (IoT) 
devices, and machine-to-machine (M2M) communication systems demands 
efficient medium access control (MAC) protocols that optimize performance 
under dynamic conditions \cite{3GPP38.321}. While reinforcement learning (RL) \cite{Naderializadeh2021Resource, Yu2019DeepRL,Ye2019V2V} can automatically 
learn optimal scheduling policies by exploring the large state-action 
space of wireless environments, a fundamental barrier to AI-native 
wireless systems is \emph{sample complexity}.  Model-free RL requires extensive 
real-world interactions to converge, and each sample incurs significant 
communication overhead in resource-constrained networks. This sample 
inefficiency makes practical deployment of learning-based MAC protocols 
infeasible, necessitating approaches that achieve fast convergence with 
minimal real-world data.

\vspace{-3mm}\subsection{Related Works}\vspace{-1mm}
In multi-agent RL (MARL)  settings \cite{feriani2021single}, distributed sensor nodes can learn decentralized policies while coordinating through control signaling with the gateway. 
In contrast to model-free RL that are sample inefficient \cite{Naderializadeh2021Resource, Yu2019DeepRL,Ye2019V2V,feriani2021single}, model-based RL (MBRL)~\cite{Park2022MBRLChannelAccess} and~\cite{Hasan2024ContinualMBRL} can significantly improve sample efficiency, requiring a small number of samples generated through real environment interactions, by learning a model of the environment dynamics and using it to generate synthetic experience for policy training. 
Complementary RL approaches include diffusion-based data augmentation for replay buffers~\cite{gao2024diffusion,Dong2024}, hierarchical MBRL with temporal abstraction~\cite{Schiewer2024}, and large language model (LLM)-based knowledge integration through reward shaping \cite{Zhang2025}. However, standard MBRL and LLM-based approaches \cite{Park2022MBRLChannelAccess,Hasan2024ContinualMBRL} often learn black-box models whose performance are tied to the training data and does not provide an understanding of which variables causally 
influence which outcomes. This opacity creates three fundamental challenges. 
First, policies learned from black-box models may exploit \emph{spurious correlations} that fail to transfer when network conditions change. Second, \emph{model errors accumulate} over planning horizons, limiting the effectiveness of long-horizon autonomous control. Third, the learned policies \emph{lack explainability}, making it impossible to verify their trustworthiness before deployment, a critical concern for mission-critical IoT systems. 
A promising approach here is to exploit causal inference \cite{thomas2024causal} to understand the underlying causal structure of wireless dynamics, enabling model-based agents to generalize reliably under distribution shifts while requiring limited real-world data.

\vspace{-2mm}\subsection{ Contributions}\vspace{-1mm}
The main contribution of this paper is a novel  MBRL for device scheduling in IoT that leverages causal graphs to explicitly represent the cause and effect relations among states, actions, observations, and rewards. By exploiting this causal structure, we aim to improve~\emph{sample efficiency} through data augmentation using more accurate environment modeling; and facilitate \emph{faster learning} via model-based rollouts using the causal structure. Our key contributions include:
\begin{itemize}
    \item We formulate the wireless MAC problem for the uplink (UL) of a multi-user IoT network using structural causal models (SCMs) that explicitly represent the causal relationships between sensor node states, control messages, transmission actions, observations, and rewards.
    \item We develop gated recurrent units (GRU) and attention-based neural networks to learn causal dependencies modeled as a SCM, with the architecture \emph{inherently providing explainability} through attention weights that reveal causal influences on scheduling decisions.
Further, we integrate the learned causal model into a model-based RL algorithm using proximal policy optimization (PPO), where the causal model generates synthetic rollouts for policy training.
\item We \emph{analytically derive the sample efficiency gains} that can be achieved using a causal MBRL compared to traditional RL baseline, as a function of the SCM complexity and learned model accuracy.
  \item Through extensive simulations\footnote{\vspace{-1mm}Due to space constraints, the evaluation of generalizability is omitted and will be presented in a journal extension.}, we show that proposed causal MBRL achieves \emph{faster convergence} compared to model-free baselines and on average \emph{better sample efficiency} via $58\%$ reduction in environment interactions.
\end{itemize}


\vspace{-2mm}\section{System Model and MARL Formulation}\vspace{-1mm}
 
We consider an IoT network with $U$ sensor nodes that communicate with a gateway over a shared UL channel. 
Each sensor node aims to transmit $P$ data packets to the gateway while coordinating channel access through control messages.
Similar to the model considered in~\cite{Valcarce2021MAC}, all sensor nodes transmit their UL data packets over the same frequency channel using time division multiple access (TDMA) scheme. The objective is 
to deliver all packets to the gateway with minimal latency, which is 
critical for time-sensitive IoT applications such as environmental 
monitoring, industrial control, and smart grid systems.
We consider a packet erasure channel in which the transmitted data packets can be received in error with block error rate (BLER) $\rho \in [0,1]$.  For simplicity, we assume the BLER is identical for all UL data transmissions. Each sensor node has a dedicated error-free UL and downlink (DL) control channels for exchanging signaling messages with the gateway. We assume that control channels do not experience packet losses and have negligible transmission delay.
The gateway implements a pre-defined MAC signaling protocol that is not learned.
Sensor nodes need to learn to communicate with the gateway by understanding the channel access scheme and MAC signaling protocol of the gateway with an objective  to successfully deliver the packets in minimum time possible.

In the UL, the gateway receives the scheduling requests (SRs) from the sensor nodes, and in the DL, it sends the scheduling grant (SG) to the specific node who has been granted the channel access.
Furthermore, in the DL, it also sends an acknowledgment (ACK) when a packet is successfully received. 
Let $\mM_L=\{0,1,2\}$ denote the set of three possible DL control messages (DCMs) from the gateway to each node and $m_t^u\in\mM_L$ denote DCM sent to $u^{\text{th}}$ node at time slot $t$. 
The gateway sending a null message is denoted by $m_t^u = 0$, releasing an SG to node $u$ in the next time slot is denoted by  $m_t^u = 1$, and sending an ACK is defined by $m_t^u = 2$. 
We define $\mathcal{U}_L = \{0,1\}$ as the set of UL control messages (UCMs), where $n_t^u \in \mathcal{U}_L$ represents the UCM at time $t$ from node $u$. The gateway will interpret  $n_t^u = 1$ as receiving  scheduling request and $n_t^u = 0$ as no request.


At each time step $t$, for each learner $u$, the environment delivers a scalar observation $o_t^u \in \mathcal{O}^U = [0, P]$
describing the number of packets that remain in the learner's UL transmit buffer. Here, $P$ is the transmit buffer capacity, and all packets are assumed to be of the same size. Similarly, at each time step, the gateway receives a scalar observation 
$
o_t^b \in \mathcal{O}^B = [0, L + 1]$
from the environment. $o_t^b = 0$ means that UL channel is idle, $o_t^b = u \in [1, U]$ represents collision-free UL transmission received from node $u$ and $o_t^b = U + 1$, when collision occurred in the UL channel.
The gateway also receives the SRs $\{n_t^u\}_{u=1}^U$ from the sensor nodes via the control channel. It selects one requesting node uniformly at random and sends an SG to that node: $m_t^u = 1$ for the selected node $u$, and $m_t^{u'} = 0$ for all other nodes $u' \neq u$. If a node successfully transmits data concurrently with sending an SR, the gateway sends an ACK instead of an SG to that node, and schedules a different node.
For $u^{th}$ sensor node at time step $t$, let $a_t^u\in\{0,1,2\}$ denotes the UL shared channel action. Action value $ 0$  correspond to idle, and action values $1$ and $2$ correspond to transmitting and deleting the oldest packet in the buffer, respectively.
We define the vector $\bmd_t^u=\left(n_t^u,a_t^u\right)$ as the combined decision made by the $u^{\textrm{th}}$ sensor node.  
Further, the state of each sensor node $\bmx_t^u$ is defined, and it incorporates the observation history, previous decisions, and received control messages over a window of $N$ time steps:
\vspace{-1mm}\begin{equation}
\begin{aligned}
      &\bmx_t^u = \left[o_{t-1}^u, \bmd_{t-1}^u,m_{t-1}^u,\cdots, o_{t-N}^u, \bmd_{t-N}^u,m_{t-N}^u\right],
\end{aligned}
\vspace{-1mm}\end{equation}
Similarly, the gateway maintains a state based on its observations and the control message history from all nodes:
\begin{equation}
\begin{aligned}    &\bmx_t^b = \left[o_{t-1}^b, \bmn_{t-1},\bmm_{t-1},\cdots, \bmo_{t-N}^b, \bmn_{t-N},\bmm_{t-N}\right],
\end{aligned}\end{equation}
where $\bmn_t$ and $\bmm_t$ are the vector of UCMs from all sensor nodes and the vector of DCMs to sensor nodes, respectively. For notational convenience, we also define the global state $\bmx_t=(\bmx_t^b,\bmx_t^1,\cdots,\bmx_t^U)$, global observation $\bmo_t=(o_t^b,o_t^1,\cdots,o_t^U)$ and global actions $\bmd_t=(\bmd_t^1,\bmd_t^2,\cdots,\bmd_t^U)$.

\vspace{-1mm}\subsection{Multi-Agent RL Framework}\vspace{-1mm}

Following \cite{Valcarce2021MAC}, we assign a reward of $r_t = -1$
once packets are successfully delivered and deleted. 
Reward is greater than $-T_{\max}$ when all packets are delivered and deleted in less number of steps.   The total episode reward 
$R \!=\! \sum\limits_{t=0}^{T-1} r_t$, 
where $T \leq T_{\max}$ is the actual episode length.  Given the current state  $\bmx_t^u$ and observation $o_t^{u}$, the RL agent at each sensor node selects shared channel action and UCM according to its policy  $\pi_u(\bmd_t^u\mid \bmx_t^u)$. The objective is to learn policies $\{\pi_u(\bmd_t^u\mid \bmx_t^u)\}_{u=1}^U$ that maximize the expected cumulative discounted reward:
$J(\pi) = \mathbb{E}_{\pi} \left[ \sum_{t=0}^{\infty} \gamma^t r_t \right]$, 
where $\gamma \in [0,1)$ is the discount factor that balances immediate and future rewards $r_t$. 
\begin{figure}
\vspace{-2mm}    \centering
    \includegraphics[width=0.5\textwidth,height=4.8cm]{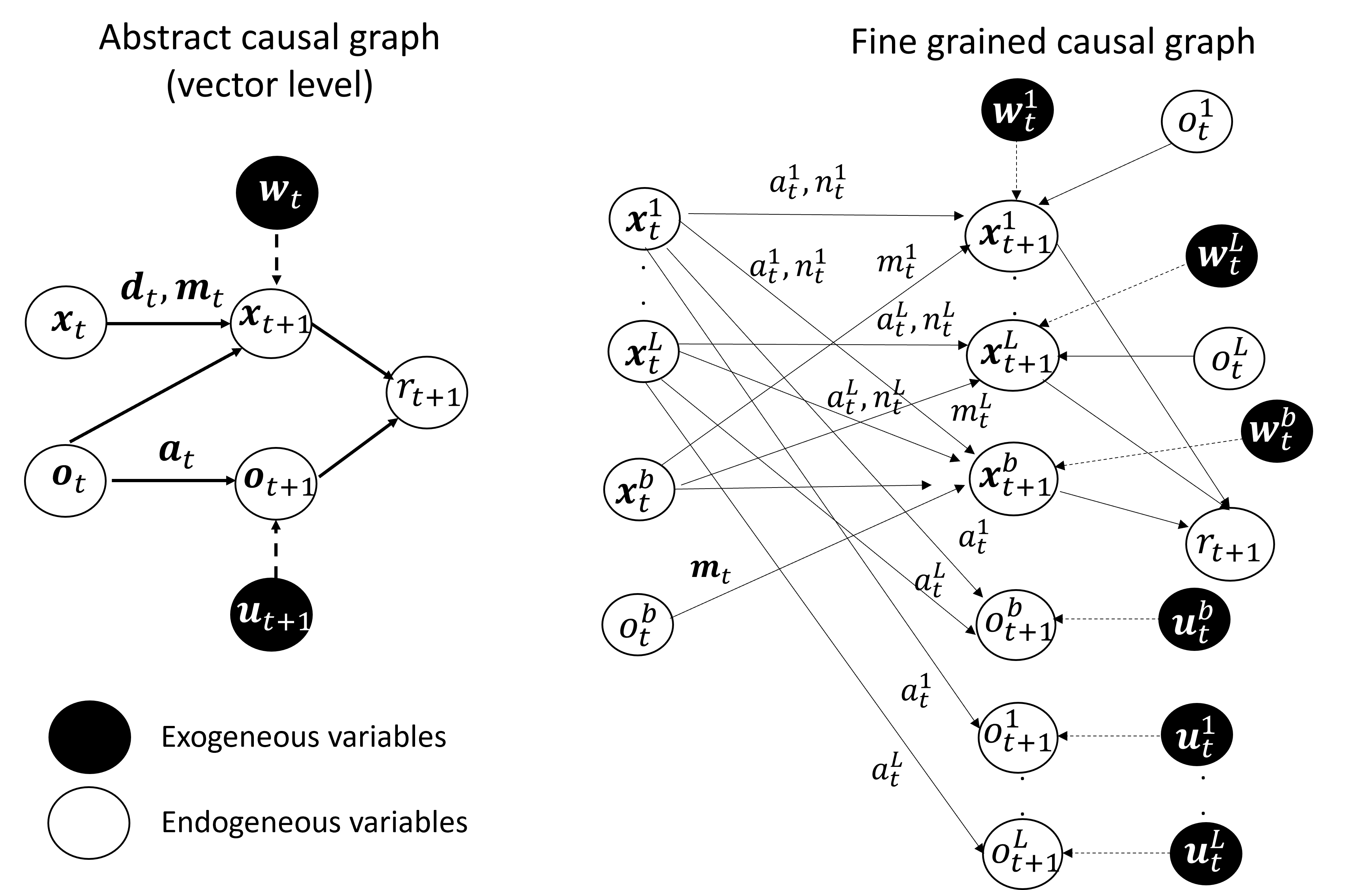}
    \vspace{-5mm}\caption{\small Causal Graph $\mG$ for MAC Scheduling. Abstract level causal graph is provided for simplicity of understanding. We use the fine grained causal graph for our proposed method.}
    \label{fig:Causal_graph}
    \vspace{-6mm}
\end{figure}
Computing node policies to maximize $J(\pi)$ using standard MARL has the limitation of requiring large number of real-world data samples, which may not be available in practice. To overcome this challenge, we now propose a causal MBRL approach that addresses these limitations via explicitly capturing the \emph{causal relations} among RL variables and enabling \emph{data augmentation} by generating synthetic data samples that  nearly matches the real-world samples. 

\vspace{-1mm}\section{Causal Model Based Wireless Scheduling}\vspace{-1mm}

In this section, we present a rigorous framework for learning causal models in multi-user IoT device scheduling for channel access. We formalize the SCM using the known causal graph $\mG$ (see Fig.~\ref{fig:Causal_graph}). 
We then formulate a causal MBRL-based scheduling problem that learns the structural equations (SEs) part of causal model and optimize the scheduling policies.

\vspace{-2mm}\subsection{Formal SCM Definition}
\vspace{-1mm}

We formalize the wireless scheduling environment as an SCM $\mathcal{M} = \langle \mathcal{U}, \mathcal{V}, \mathcal{F}, P(\mathcal{U}) \rangle$, where $\mathcal{U}$ is the set of exogenous variables (external noise and unobserved factors),  $\mathcal{V}$ is the set of endogenous variables (system variables determined by the model), $\mathcal{F}$ is the set of SEs defining causal relationships, and $P(\mathcal{U})$ is the probability distribution over exogenous variables.  The endogenous variables at time $t$, defined as $\mathcal{V}_t$, consist of agent states $\mathcal{X}_t = \{\bmx_t^1, \ldots, \bmx_t^U, \bmx_t^b\}$, observations $\mathcal{O}_t = \{o_t^1, \ldots, o_t^U, o_t^b\}$, UL actions $\mathcal{D}_t = \{\bmd_t^1, \ldots, \bmd_t^U\}$ (sensor node decisions), DCMs $\mathcal{M}_t = \{m_t^1, \ldots, m_t^U\}$ (gateway decisions) and reward $r_t \in \mathbb{R}$. The exogenous variables capture stochasticity and include state transition noise $ \{w_t^{1}, \ldots, w_t^{L}, w_t^{b}\}$ (e.g., due to channel fading, interference), and external environmental factors $ \{u_t^1, \ldots, u_t^U\}$ . Each $u_t^u$ corresponds to the packet arrival process that generate the observations $o_t^u$. Each endogenous variable $v \in \mathcal{V}_t$ is defined by a SE: $
    v := f_v(\text{Pa}(v), \mathcal{U}_v),$
where $\text{Pa}(v) \subseteq \mathcal{V}_{t-1} \cup \mathcal{V}_t$ are the parent variables of $v$, i.e., all variables that directly cause $v$, in the causal graph, and $\mathcal{U}_v$ represents the exogenous inputs affecting $v$.  Specifically, we have \emph{state transition equations} as
\vspace{-2mm}\begin{equation}
\begin{aligned}
    \bmx_{t+1}^u &:= f_x^u(\bmx_t^u, o_t^u, \bmd_t^u, m_t^u, w_t^{u}), \quad u = 1, \ldots, U,\\
\bmx_{t+1}^b &:= f_x^b(\bmx_t^b, o_t^b, \boldsymbol{d}_t, \boldsymbol{m}_t, w_t^{b}),
\vspace{-2mm}\label{eq_se_1}
\end{aligned}
\vspace{-2mm}\end{equation}
and \emph{observation equations} as
\vspace{-2mm}\begin{equation}
\begin{aligned}
o_{t+1}^u &:= f_o^u(\bmx_t^u, a_t^u, u_t^u), \quad u = 1, \ldots, U,\\
o_{t+1}^b &:= f_o^b(\bmx_t^b, \boldsymbol{d}_t, w_t^b),  
\label{eq_se_2}
\vspace{-2mm}\end{aligned}\vspace{-2mm}\end{equation}
\vspace{-2mm}\begin{equation}\hspace{-13mm}\mbox{and \emph{reward equation} as}\,\,\, r_{t+1} := f_r(\boldsymbol{x}_t, \boldsymbol{d}_t, \boldsymbol{m}_t, \boldsymbol{o}_t).\label{eq_se_3}
\vspace{-2mm}\end{equation}
The causal relationships in the SCM are represented by a directed acyclic graph (DAG) $\mathcal{G} = (\mathcal{V}, \mathcal{E})$, where nodes correspond to endogenous variables $\mathcal{V} = \mathcal{V}_t \cup \mathcal{V}_{t+1}$
 and directed edges represent direct causal relationships: $(v_i, v_j) \in \mathcal{E}$ if $v_i \in \text{Pa}(v_j)$.
Unlike 
knowledge graphs that represent semantic relationships or correlative 
associations \cite{louizos2017causal}, SCM is specifically designed to 
capture \emph{direct causal influences} and avoids the exponential blowup 
of learning unstructured dependencies. It preserves 
\emph{interventional} semantics, i.e., the graph remains valid when 
network conditions vary, enabling policies  generalize across channel conditions. SCMs also satisfy the Markov property, i.e., each variable is conditionally independent of its non-descendants given its parents.

\vspace{-1mm}\subsection{Problem Formulation: Causal MBRL Based Scheduling }\vspace{-1mm}
Given $\mathcal{G}$, we learn the functional forms of the SEs in \eqref{eq_se_1}-\eqref{eq_se_3} using parameterized inference networks.\footnote{\scriptsize We note that constraint-based methods (PC algorithm) or score-based approaches (greedy equivalence search, NOTEARS) could be employed in future work to infer $\mathcal{G}$ from wireless data.} 
For each endogenous variable $v_j \in \mathcal{V}$, the objective is to learn a conditional distribution: $
    q_\phi(v_j \mid \text{Pa}(v_j)) \approx p(v_j \mid \text{Pa}(v_j))$, by minimizing the negative log-likelihood $-\sum_{v_j\in \mV_{t+1}} \mathbb{E}_{\delta \sim \mathcal{D}} [\log q_\phi(v_j | \text{Pa}(v_j))]$,
where $q_\phi$ is parameterized by neural network weights $\phi$, and $p$ represents the true conditional distribution in the environment. Here, the expectation is over the empirical distribution $p(v_j \mid \text{Pa}(v_j))$ obtained using the wireless data $\mD$ that involves the stat-action-reward trajectories. Challenges here involve heterogeneous parent variable types and dimensions across $\text{Pa}(v_j)$, dynamic structural changes as the causal graph evolves over time, and complex temporal dependencies among sequential actions. To handle heterogeneous inputs, we introduce a \emph{variable encoding layer} that maps all variables to a common embedding space.
This variable encoder is a two-layer MLP with ReLU activation function.

Given the current-time variables
$\mathcal{V}_t$, we obtain the encoded set 
$\hat{\mathcal{V}}_t = \{\hat{s}_i = f(s_i) : s_i \in \mathcal{V}_t\}$. Encoders are shared across all inference networks, enabling transfer learning. The architecture for inference network is inspired from the causal model discussed in \cite{YuArxiv2023}. For each target variable $v_j \in \mathcal{V}$, we construct an inference network $f^j: \text{Pa}(v_j) \rightarrow \mathcal{P}(v_j)$ that outputs a probability distribution over $v_j$.
\begin{figure*}
\begin{equation}
    f^j: \text{Pa}(v_j) \xrightarrow{\text{encode}} \text{Pa}(v_j)_{\text{encoded}} \xrightarrow{\text{separate}} \begin{cases} \text{States } \mathcal{S}_j \\ \text{Actions } \mathcal{A}_j \end{cases} \xrightarrow{\text{process}} \boldsymbol{h}^j \xrightarrow{\text{decode}} q_\phi(v_j \mid \text{Pa}(v_j))   
.\label{eq_architecture}
\end{equation}
\vspace{-6mm}\end{figure*}
We partition the parent set into state variables and action variables:
$\text{Pa}(v_j) = \mathcal{S}_j \cup \mathcal{A}_j$,
where $\mathcal{S}_j = \text{Pa}(v_j) \cap (\mathcal{X}_t \cup \mathcal{O}_t)$ are the states and observations and $\mathcal{A}_j = \text{Pa}(v_j) \cap (\mathcal{D}_t \cup \mathcal{M}_t)$ are the actions and control messages. This partitioning is considered because states are processed independently, while actions are processed sequentially to capture temporal dependencies. Further, the integrated causal MBRL problem aims at optimizing the causal model neural network parameters $\phi$ and policy model parameters $\theta$:
\vspace{-5mm}\begin{equation}
\min_{\theta, \phi} \sum_{i=1}^{n_{\text{epoch}}} \left[ \mL_{\text{RL}}^{(i )}(\theta, \phi) + \mathcal{L}_{\text{model}}^{(i)}(\phi) \right],\label{eq_Problem}
\vspace{-4mm}\end{equation}
where the RL loss  $
\mL_{\text{RL}}^{(i)}(\theta, \phi) = -\mathbb{E}_{\pi_\theta, q_\phi} \left[ \sum\limits_{t=0}^{T} \gamma^t r_t \right]$ is averaged over the real-world interaction with the wireless environment,  $n_{\text{epoch}}$ is the number of real-world data interactions stored in $\mD$,
and the causal model loss is
\begin{equation}
\mathcal{L}_{\text{model}}^{(i)}(\phi) = 
\!-\!\!\sum_{v_j\in \mV_{t+1}} \!\!\mathbb{E}_{\delta \sim \mathcal{D}^{(i)}} [\log q_\phi(v_j | \text{Pa}(v_j))] + \lambda \|\phi\|_2^2.\nonumber
\vspace{-1mm}\end{equation}
For a batch $\mathcal{B} \subset \mathcal{D}$ of size $B$, $\mathcal{L}_{\text{model}}(\phi) = -\frac{1}{B} \sum_{\delta \in \mathcal{B}} \sum_{v_j \in \mathcal{V}_{t+1}} \log q_\phi(v_j^\delta \mid \text{Pa}(v_j)^\delta),$
where $v_j^\delta$ is the observed value of $v_j$ in transition $\delta$.

\section{Proposed Solution: Causal MBRL Algorithm}
We now describe the MBRL algorithm and analyze sample efficiency gains.
We propose to solve $\phi$ and $\theta$ in an alternating way via a three-phase approach executed 
each epoch to optimize \eqref{eq_Problem}: (i) collect real environment trajectories using $\pi_\theta$, (ii) periodically 
update the causal model parameters $\phi$ using accumulated data, and (iii) optimize the policy network parameters $\theta$ 
using both real and synthetic data generated from the learned model. The detailed steps are provided in Algorithm~\ref{alg:model_based_rl}. We model the causal inference network using attention based neural networks.

\vspace{-2mm}\subsection{Causal Inference Networks}
 An overview of the architecture is represented by \eqref{eq_architecture}.
\subsubsection{Value vector computation}
For each state variable $s_i \in \mathcal{S}_j$, we compute a contribution vector (aka value vector) 
$\boldsymbol{c}_i^j$ using a single layer MLP.  $\boldsymbol{c}_i^j$ represents the potential contribution of state $s_i$ to predicting $v_j$. 
\subsubsection{Action embedding with GRU}
Given encoded actions $\{\hat{a}_i\}_{a_i \in \text{PA}(v_j)}$, the GRU computes the action embedding $\boldsymbol{e}^j = \text{GRU}^j(\{\hat{a}_i\})$. From this, we derive the query vector $\boldsymbol{q}^j = W_q^j \boldsymbol{e}^j + \boldsymbol{b}_q^j$ and action contribution vector $\boldsymbol{c}_a^j = W_a^j \boldsymbol{e}^j + \boldsymbol{b}_a^j$, where weights $W_q^j, W_a^j$ and biases are learnable parameters.
Each state variable $s_i$ has a learnable key vector $\boldsymbol{k}_i$ (shared across all inference networks to ensure consistent variable importance representations). The attention score $e_i^j = \boldsymbol{k}_i^T \boldsymbol{q}^j$ measures compatibility between state $s_i$ and query $\boldsymbol{q}^j$. We normalize attention weights using:
\vspace{-2mm}\begin{align}
    \alpha_i^j &= \frac{\exp(e_i^j)}{1 + \sum_{s_k \in \mathcal{S}_j} \exp(e_k^j)}, \,\,\alpha_a^j &= \frac{1}{1 + \sum_{s_k \in \mathcal{S}_j} \exp(e_k^j)}.\nonumber\vspace{-2mm}
\end{align}
The attention weights provide interpretability of the learned causal model.  High attention weight $\alpha_i^j$ indicates that state $s_i$ has strong causal influence on $v_j$ given the current action context. We then compute the hidden representation of the posterior distribution as a weighted sum of the value vectors:
\vspace{-2mm}\begin{equation}
    \bmh^j = \sum_{s_i \in PA(v_j)} \alpha_i^j \cdot \bmc_i^j + \alpha_a^j \cdot \bmc_a^j.
\vspace{-1mm}\end{equation}
The final component maps the hidden representation to distribution parameters.
The decoder $D^j: \mathbb{R}^{d_{h}} \rightarrow \Theta_{v_j}$ is a two-layer MLP, 
 where $\Theta_{v_j}$ is the parameter space for the distribution of $v_j$. For continuous variables (e.g., states, observations):
$q_\phi(v_j \mid \text{Pa}(v_j)) = \mathcal{N}(\mu_j, \sigma_j^2)$,
where $[\mu_j, \log \sigma_j] = D^j(\boldsymbol{h}^j)$. We predict $\log \sigma_j$ to ensure positivity.
For discrete variables (e.g., discrete actions):
$q_\phi(v_j \mid \text{Pa}(v_j)) = \text{Categorical}(\boldsymbol{\pi}_j)$
where $\boldsymbol{\pi}_j = \text{softmax}(D^j(\boldsymbol{h}^j))$.

\subsubsection{Policy update using data augmentation} We now integrate the learned causal model into a complete MBRL framework.
\setlength{\textfloatsep}{0pt}\begin{algorithm}[t]
\caption{Proposed Causal Model-Based RL}
\label{alg:model_based_rl}\vspace{-0.1cm}
\begin{algorithmic}[1]\small
\STATE \textbf{Input:} Environment with unknown dynamics; initial policy parameters $\theta_{\text{policy}}$; hyperparameters $n_{\text{epoch}}, n_{\text{graph}}, n_{\text{round}}, k_{\text{rollout}}$
\STATE \textbf{Output:} Optimized policy $\pi_\theta$
\STATE \textbf{Initialization:} Replay buffer $\mathcal{D} \leftarrow \emptyset$, causal graph $\mathcal{G}$ as in Fig.~\ref{fig:Causal_graph}, inference network parameters $\phi \!\sim\! \mathcal{N}(0, 0.01)$, policy networks $\theta_{\text{policy}} \!\sim \!\mathcal{N}(0, 0.01)$, value network $\theta_{\text{value}} \!\sim \!\mathcal{N}(0, 0.01)$.
\FOR{$i = 1$ to $n_{\text{epoch}}$}
    \STATE \textbf{Phase 1: Real Environment Interaction}
    \FOR{each episode $e = 1$ to $n_{\text{episodes}}$}
        \STATE Initialize state $\bmx_0 \sim p_0(\bmx)$
        \FOR{$t = 0$ to $T-1$}
            \STATE Sample actions: $\bmd_t^u \sim \pi_u(\cdot | \bmx_t^u)$ $\forall u $; $\bmm_t \sim \pi_b(\cdot | \bmx_t^b)$
            \STATE Execute actions; observe $\bmx_{t+1}, \bmo_{t+1}, r_t$, store in $\mathcal{D}$
        \ENDFOR
    \ENDFOR
    \STATE \textbf{Phase 2: Causal Model Update}
    \IF{$i \bmod n_{\text{graph}} = 0$}
        \FOR{gradient step $= 1$ to $N_{\text{model}}$}
            \STATE Sample minibatch $B \subset \mathcal{D}$
            \STATE Compute  $\mL_{\text{model}}(\phi)$, and  update $\phi \leftarrow \phi - \eta_{\text{model}} \nabla_\phi \mL_{\text{model}}$
        \ENDFOR
    \ENDIF
    \STATE \textbf{Phase 3: Policy Optimization}
    \FOR{$j = 1$ to $n_{\text{round}}$}
        \STATE Generate synthetic data: $\mathcal{D}_{\text{sim}} \leftarrow \emptyset$
        \STATE Sample initial states $\{x_0^{(i)}\}_{i=1}^{N_{\text{rollout}}}$ from $\mathcal{D}$
        \FOR{$i = 1$ to $N_{\text{rollout}}$}
            \STATE trajectory \!$\leftarrow\! k_{\text{rollout}}(\bmx_0^{(i)}\!, \pi, k_{\text{rollout}})$ using Algorithm 2
            \STATE $\mathcal{D}_{\text{sim}} \leftarrow \mathcal{D}_{\text{sim}} \cup$ trajectory
        \ENDFOR
        \STATE $\mathcal{D}_{\text{combined}} \leftarrow \mathcal{D} \cup \mathcal{D}_{\text{sim}}$
        \FOR{ppo\_epoch $= 1$ to $N_{\text{ppo}}$}
            \STATE Sample minibatch $B \subset \mathcal{D}_{\text{combined}}$, Compute  $\{\hat{A}_t\}$
            \STATE Update policy: $\theta_{\text{policy}} \leftarrow \theta_{\text{policy}} - \eta_{\text{policy}} \nabla_\theta \mL_{\text{PPO}}$
            \STATE Update value: $\theta_{\text{value}} \leftarrow \theta_{\text{value}} - \eta_{\text{value}} \nabla_\theta \mL_{\text{value}}$
        \ENDFOR
    \ENDFOR
\ENDFOR
\STATE \textbf{Return:} Learned policy $\pi_\theta$
\end{algorithmic}
\end{algorithm}
\setlength{\textfloatsep}{0pt}\begin{algorithm}[t]
\caption{$k$-Step Model Rollout}\vspace{-0.1cm}
\label{alg:step_model}
\begin{algorithmic}[1]\small
\STATE Initialize $\mD_{\text{sim}}\leftarrow []$, $\bmx \leftarrow \bmx_0$
\FOR{$\tau = 0$ to $k-1$}
     \STATE $\bmd_{\tau}^u \sim \pi_u(\cdot \mid \bmx_{\tau}^u), \forall u$.     \text{// Sample actions from current policy}:
    \STATE $\bmm_{\tau} \sim \pi_b(\cdot \mid \bmx_{\tau}^b)$
    
    \text{// Predict next state and observations using causal model}
    \FOR{$u = 1$ to $U$}
        \STATE $\bmx_{\tau+1}^u \!\sim\! q_{\phi}\bm(\bmx_{\tau+1}^u \!\mid\! \text{Pa}(\bmx_{\tau+1}^u))$, $o_{\tau+1}^u \!\sim\! q_{\phi}(o_{\tau+1}^u \!\mid\! \text{Pa}(o_{\tau+1}^u))$
    \ENDFOR
    \STATE 
$r_{\tau} = f_r(\bmx_{\tau}, \bmd_{\tau}, \bmm_{\tau}, \bmo_{\tau})$ \text{// Compute reward}
    \STATE 
    $\mD_{\text{sim}} \leftarrow \mD_{\text{sim}} \cup \{(\bmx_{\tau}, \bmd_{\tau}, \bmm_{\tau}, r_{\tau}, \bmx_{\tau+1}, \bmo_{\tau+1})\}$
    \STATE  
  $\bmx \leftarrow \bmx_{\tau+1}$ \text{// Update current state}
\ENDFOR
\RETURN $\mD_{\text{sim}}$
\end{algorithmic}
\end{algorithm}
We leverage the learned causal model to generate synthetic data matching real-world dynamics, reducing the need for extensive wireless samples during training. Short rollouts ($k \in [3,10]$) for synthetic data reduce model error accumulation for higher accuracy but sacrifice sample efficiency, whereas long rollouts $k \in [10,50]$ improve efficiency at the cost of compounding errors.  Further, PPO is used to optimize the network parameters because it safely handles training on both real and simulated data through importance weighting and clipping, which is critical for MBRL.
Using both real and simulated data, our loss function $\mathcal{L}^{\text{PPO}}(\theta)$  is equal to
\vspace{-1mm}
\begin{equation}
\mathbb{E}_{_{(\bmx_t,\bma_t) \sim D}}\!\!\left[
\min\left(
\rho_t(\theta),\,
\text{clip}\big(\rho_t(\theta), 1-\epsilon, 1+\epsilon\big)
\right)\hat{A}_t
\right]
\end{equation}
where
$
\rho_t(\theta) \!\!=\!\! \frac{\pi_\theta(\bmd_t \mid \bmx_t)}{\pi_{\theta_{\text{old}}}(\bmd_t \mid \bmx_t)},$ $
\hat{A}_t    $ is the advantage estimate.
The advantage function is formally defined as $
    A^\pi(\bmx_t, \bmd_t) \!=\! Q^\pi(\bmx_t, \bmd_t) - V^\pi(\bmx_t),$
where $Q^\pi(\bmx_t, \bmd_t)$ is the expected return when taking action $\bmd_t$ in state $\bmx_t$ and then following policy $\pi$, and $V^\pi(\bmx_t)$ is the expected return in state $\bmx_t$ and following policy $\pi$.
$\textrm{clip}()$ function constraints $\rho_t$ within $\{1-\epsilon, 1+\epsilon\}$.

\vspace{-2mm}\subsection{Theoretical Analysis}

We justify sample efficiency through causal factorization. Under standard assumptions 
(identifiability, bounded state, and action spaces, Markov property, Lipschitz dynamics), the sample 
complexity for causal MBRL is:
\vspace{-2mm}\begin{theorem}
\label{thm:sample_complexity}
For a wireless MAC system with $\abs{\mathcal{V}}$ variables (includes state, action and reward), maximum support $V_{\max}$, 
causal in-degree $d_{\text{in}}$ (maximum number of parents for a single node in $\mG$), and a  
 desired accuracy of the causal model $\epsilon > 0$,  causal MBRL achieves sample complexity:
\vspace{-0mm}\begin{equation}
N_{\text{causal}} = {\mathcal{O}}\left({|V|^2 \cdot V_{\max}^{d_{\text{in}}}}/{\left(\epsilon^2(1-\gamma)\right)}\right)
\vspace{-1mm}\end{equation}
versus black-box MBRL's ${\mathcal{O}}(V_{\max}^{|V|}/\epsilon^2)$, 
yielding exponential improvement $\sim V_{\max}^{\abs{\mathcal{V}}-d_{\text{in}}}/\abs{\mathcal{V}}^2$. 
\vspace{-1mm}\end{theorem}
\vspace{-1mm}\begin{IEEEproof} Detailed proof is omitted due to space constraints. The key insight is that the joint distribution over all variables can be factorized using causal structure $
p(\mathcal{V}) = \prod\limits_{v_j \in \mathcal{V}} p(v_j | \text{Pa}(v_j))$.
Each conditional $p(v_j | \text{Pa}(v_j))$ depends only on $d_{\text{in}}(v_j)$ parents, where $d_{\text{in}}(v_j) \leq d_{\text{in}} \ll |\mathcal{V}|$ by the acyclic structure constraint.
By standard results in learning theory (Hoeffding's inequality and finite hypothesis classes) ~\cite{Hoeffding1963,ShalevShwartz2014}, to learn a single conditional $p(v_j | \text{Pa}(v_j))$ with accuracy $\epsilon$, we require samples proportional to the support size:
$N_j = \mathcal{O} \left( {V_{\max}^{d_{\text{in}}(v_j)}}/{\epsilon^2} \right)$.
This is because the conditional probability table has $\mathcal{V}_{\max}^{d_{\text{in}}}$ entries (one for each parent combination), each requiring $\mathcal{O}(\log(1/\epsilon)/\epsilon^2)$ samples for accurate estimation.
Since we must learn all $|\mathcal{V}|$ conditionals, and learning one variable does not immediately improve learning others (by causality), the total complexity is:
\vspace{-1mm}\begin{equation}
N_{\text{total}} = \sum_{v_j \in \mathcal{V}} N_j = |\mathcal{V}| \cdot \mathcal{O}\left(\frac{\mathcal{V}_{\max}^{d_{\text{in}}}}{\epsilon^2}\right) = \mathcal{O}\left(\frac{|\mathcal{V}| \cdot \mathcal{V}_{\max}^{d_{\text{in}}}}{\epsilon^2}\right)\nonumber
\vspace{-1mm}\end{equation}
The $|\mathcal{V}|^2$ factor appears when accounting for variance reduction across epochs and confidence bound union over variables (Hoeffding-style bounds) \cite{boucheron2003concentration}.
\end{IEEEproof}
By leveraging causal structure, Theorem~\ref{thm:sample_complexity} shows that learning complexity scales with the in-degree $d_{\text{in}}$ rather than the total number of variables $\abs{\mathcal{V}}$; since wireless dynamics are sparse (few variables directly influence each outcome), this causal factorization dramatically reduces sample requirements compared to unstructured black-box approaches.

\section{Simulation Results and Analysis}
\label{Simulation}

We validate the sample efficiency and performance of the proposed causal MBRL based wireless scheduling using extensive Monte-Carlo simulations.  
We implemented and tested the training procedure described in Algorithms~\ref{alg:model_based_rl} and~\ref{alg:step_model}. We have collected the performance, i.e., total reward $R$, per episode.
We start our simulations with a full buffer at the user and repeated the training for maximum timesteps per episode $T_{\max}=4, 8, 16, 32$.
The dimension of the variable encoder embedding vector is chosen to be $d = 64$. For the inference networks, the input dimension is  $64$ and the dimension of one hidden layer $d_h = 128$. The dimension of the query vector is $d_q = 32$, and the ReLU activation function is used. 
The output dimension is $2$ representing mean and variance of the causal node with softmax as the activation function and cross-entropy loss function. The policy network is similar with input dimension $64$, one hidden layer with dimension equal to $128$, and $tanh$ is used as the activation function. The output dimension is $3$, outputting the UL shared channel action, UCM, and the logarithm of the sum of probabilities of channel and UL actions. Cross-entropy loss function is employed. Training parameters are listed in Table~\ref{table:training_hyperparameters}.

{\em Pre-defined policy at the node:}
We compare the reward of our proposed causal MBRL with the one obtained when the nodes implement a pre-defined MAC signaling policy known to the gateway as
described in Algorithms~1 and~2 of~\cite{Valcarce2021MAC}.
{\em Tabular Q-learner:} We also compare with ~\cite{Valcarce2021MAC} that chooses the action and uplink signaling based on Q-tables.



\begin{table}[t]
\centering\scriptsize
\vspace{-1mm}
\caption{Hyperparameters and algorithm configuration}
\label{table:training_hyperparameters}
\begin{tabular}{lrlr}
\hline
\textbf{Hyperparameter} & \textbf{Value} & \textbf{Hyperparameter} & \textbf{Value} \\
\hline
Learning rate $\alpha$ & $1e{-4}$ & Discount factor $\gamma$ & $0.99$ \\
GAE parameter $\lambda$ & $0.95$ & Rollout horizon $k_{\text{rollout}}$ & $6$ \\
$N_{\text{rollout}}$ & $50$ & $N_{\text{model}}$ & $100$ \\
$N_{\text{PPO}}$ & $5$ & Batch size & $64$ \\
Episodes per epoch & $20$ & $n_{\text{graph}}$ & $1$ \\
$n_{\text{round}}$ & $4$ & Number of evaluation episodes & $128$ \\
History $N$ ($U=2$) & $3$ & History $N$ ($U=1$) & $1$  \\
\hline
\end{tabular}
\end{table}

\begin{figure}[t]
\centering
\begin{subfigure}[b]{0.49\columnwidth}
    \centering    \includegraphics[width=\columnwidth]{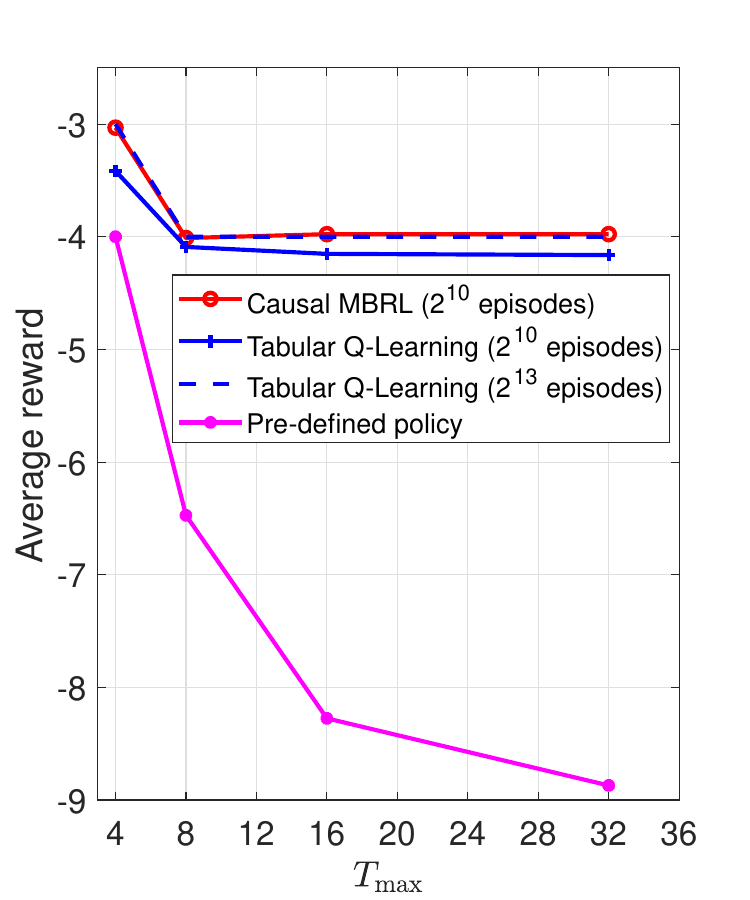}
   \vspace{-6mm} 
   \caption{\small $U=1$}
\label{fig:rewards_vs_Tmax_method1}
\end{subfigure}
\hfill
\begin{subfigure}[b]{0.49\columnwidth}
    \centering
    \includegraphics[width=\columnwidth]{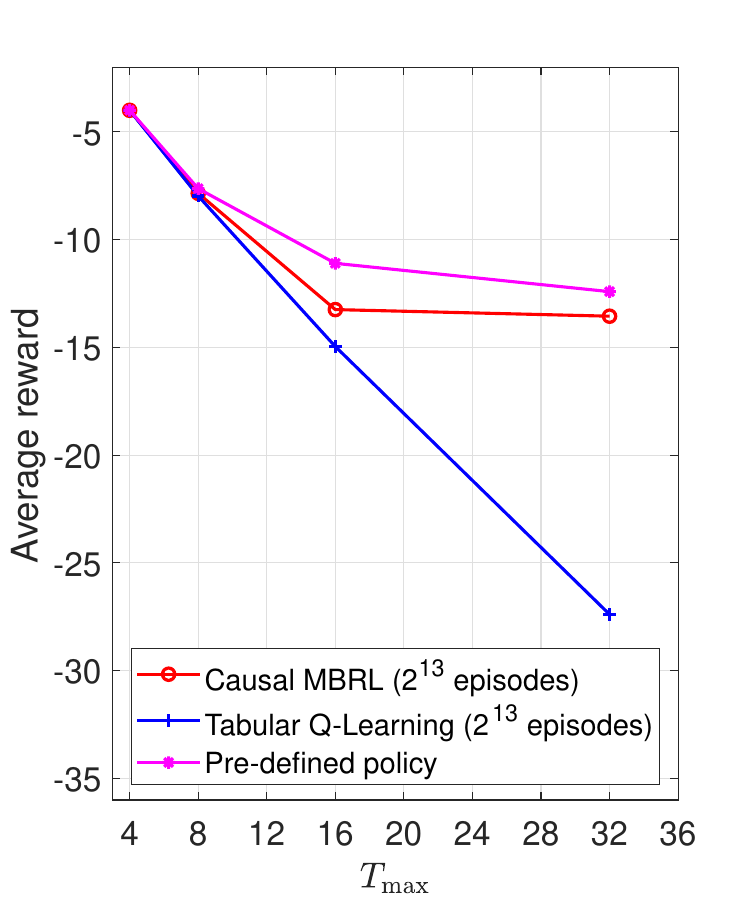}
\vspace{-6mm}\caption{\small $U=2$}
    \label{fig:rewards_vs_Tmax_method2}
\end{subfigure}
\vspace{-7mm}\caption{\small Causal MBRL wireless scheduling: Average episode rewards vs $T_{\text{max}}$ showing performance across different $U$ (BLER= 0.5). }
\label{fig:rewards_vs_Tmax}
\vspace{-1mm}\end{figure}

{\em Sample efficiency analysis:} 
Figs.~\ref{fig:rewards_vs_Tmax_method1} and~\ref{fig:rewards_vs_Tmax_method2} shows the average reward as a function of the maximum timesteps per episode $T_{\max}$ for $U=1$ and $U=2$, respectively. They compare the performance of the proposed causal MBRL, the tabular Q-learning approach~\cite{Valcarce2021MAC}, and the pre-defined policy. 
MBRL training is done for $2^{10}$ and $2^{13}$ episodes for $U=1$ and $U=2$, respectively. Furthermore, the reward shown is averaged over $8$ random seeds. 
For {\em $U=1$}, we see that after $2^{10}$ episodes, the performance of the causal MBRL converges to higher value than tabular Q-learning due to the synthetic rollouts from the learned causal model. 
Furthermore, tabular Q-learning requires $2^{13}$ episodes to converge, in turn, requiring a higher number of samples. 
Pre-defined policy performs worse because it waits for the schedule grant, unlike the learning approaches.
For {\em $U=2$}, we have shown the performance for $2^{13}$ episodes.  
Here, the causal MBRL is significantly better than Q-learning. 
For $T_{\max}= 32$, the reward is $-13.5$ and     $-27$ for the causal MBRL and Q-learning, respectively. 
Here, the pre-defined policy seems to perform better as more number of training episodes are required for learning approaches to converge.

Table~\ref{table_sampleefficiency} compares the number of samples required. It shows that the number of real-world samples required until convergence for causal MBRL is $66.66\%$ and $47.9\%$ (and an average $58\%$) less than that of tabular Q-learning for $U=1$ and $U=2$, respectively. This shows the superior sample efficiency of the proposed causal MBRL method.

\begin{table}[t]\small
\centering
\vspace{-1mm}\caption{Comparison of number of samples required}
\label{table:sample_efficiency}
\begin{tabular}{lcccc}
\hline
{Nodes} & {Q-Learning} & {Causal MBRL }\! & {Efficiency} \\
\hline
{$U=1$} & $37.5$K  &$12.5$K  & $66.66\%$ \\
\hline
\!\!{$U=2$}\!\!\!\! & $243.1$K   & $126.7$K  & $47.9\%$\\
\hline
\end{tabular}
\begin{flushleft}
\footnotesize
\small All values represent total environment interactions (real-world samples) 
required to reach $95\%$ optimal reward. 
\end{flushleft}
\label{table_sampleefficiency}
\vspace{-1mm}\end{table}


\begin{figure}[t]
\centering
\begin{subfigure}[b]{0.49\columnwidth}
    \centering    \includegraphics[width=\columnwidth]{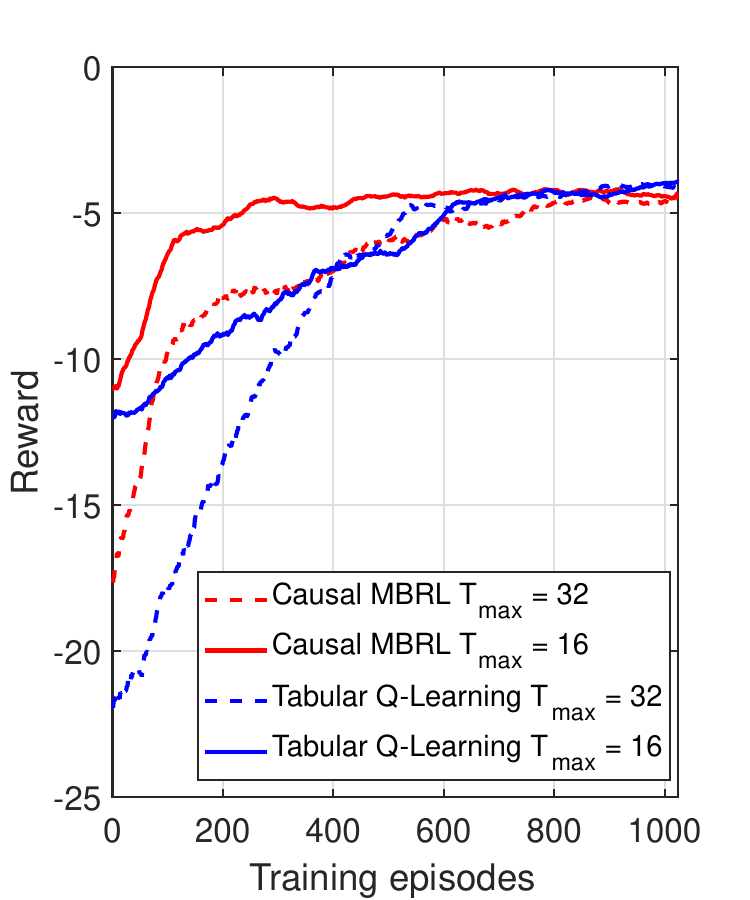}
   \vspace{-6mm} \caption{\small $U=1$}
\label{fig:rewards_vs_epi_U_1}
\end{subfigure}
\hfill
\begin{subfigure}[b]{0.49\columnwidth}
    \centering
    \includegraphics[width=\columnwidth]{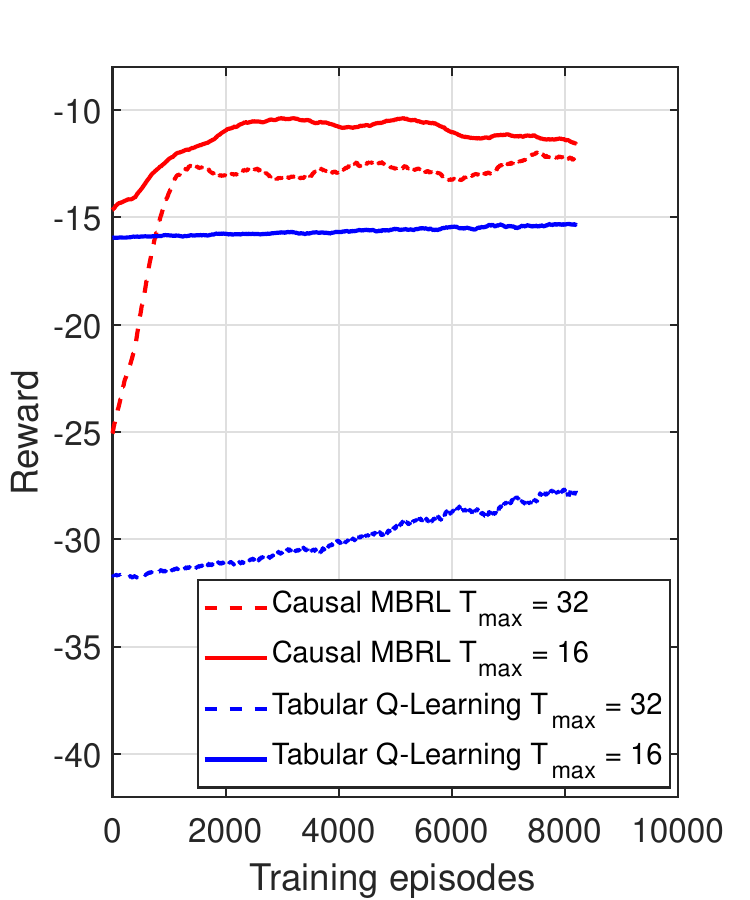}
\vspace{-6mm}\caption{\small $U=2$}
    \label{fig:rewards_vs_epi_U_2}
\end{subfigure}
\vspace{-7mm}\caption{\small  Average rewards vs training episodes for different $T_{\max}$.}
\label{fig:rewards_vs_episodes}
\end{figure}

Fig.~\ref{fig:rewards_vs_episodes} shows average reward as a function of the training episodes for the causal MBRL and tabular Q-learning. This is done for $T_{\max}=16$ and $T_{\max}=32$.   
We see that the reward increases and then saturates for both the learning approaches. For the smaller number of episodes, the increase is higher for the causal MBRL compared to the tabular Q-learning, which shows that our approach learns faster with fewer samples. 
We see that the reward of the causal MBRL saturates faster for $T_{\max}=16$ compared to $T_{\max}=32$ as the number of steps in each episode is smaller.
However, tabular Q-learning takes around the same number of episodes to saturate.

\vspace{-1mm}\section{Conclusion}\vspace{-1mm}
We have introduced causal models for IoT device scheduling, addressing 
fundamental limitations of existing MBRL 
approaches. Unlike black-box neural network models, our method discovers 
sparse DAGs that explicitly represent causal dependencies 
between states, actions, and rewards. The resulting causal structure enables 
(i) improved sample efficiency through reduced parameterization, (ii) better 
generalization via transferable causal mechanisms, and (iii) interpretability 
through attention-based causal attribution. 
Our attention-based inference networks learn to understand unknown MAC signaling protocols while coordinating channel access across multiple agents. Our simulations validate the approach and demonstrate significant improvements in convergence speed, throughput, and collision avoidance compared to model-free baselines. 
\vspace{-2mm}\bibliographystyle{IEEEbib}
\def\baselinestretch{0.9}
\vspace{-1mm}
\bibliography{refs,semantics}
\end{document}